\documentclass[sigconf]{acmart}
\AtBeginDocument{%
  \providecommand\BibTeX{{%
    \normalfont B\kern-0.5em{\scshape i\kern-0.25em b}\kern-0.8em\TeX}}}

\copyrightyear{2024}
\acmYear{2024}
\setcopyright{acmlicensed}\acmConference[UIST '24]{The 37th Annual ACM Symposium on User Interface Software and Technology}{October 13--16, 2024}{Pittsburgh, PA, USA}
\acmBooktitle{The 37th Annual ACM Symposium on User Interface Software and Technology (UIST '24), October 13--16, 2024, Pittsburgh, PA, USA}
\acmDOI{10.1145/3654777.3676341}
\acmISBN{979-8-4007-0628-8/24/10}

\usepackage{color,soul}
\usepackage{booktabs} 
\usepackage{subfig} 
\usepackage{graphicx}

\usepackage[utf8]{inputenc}
\usepackage[T1]{fontenc}

\usepackage{siunitx}
\newcommand{\hide}[1]{}

\sethlcolor{yellow}
\ifdefined\scifihidecomments
    \newcommand{\cz}[1] {}
    \newcommand{\hyunc}[1] {} 
    \newcommand{\yax}[1] {} 
    \newcommand{\sony}[1] {} 
    \newcommand{\ReviewerFeedback}[1] {}  
    \newcommand{\fran}[1] {}
    \newcommand{\rz}[1]{}
    \newcommand{\lw}[1] {}
    \newcommand{\mose}[1] {} 
    \newcommand{\ke}[1] {} 

\else
    \definecolor{burntorange}{rgb}{0.8, 0.33, 0.0}
    \definecolor{cadmiumgreen}{rgb}{0.0, 0.42, 0.24}
    \definecolor{cobalt}{rgb}{0.0, 0.28, 0.67}
    \definecolor{amber}{rgb}{1.0, 0.75, 0.0}
    \definecolor{fashionfuchsia}{rgb}{0.96, 0.0, 0.63}
    \definecolor{brightcerulean}{rgb}{0.11, 0.67, 0.84}
    \definecolor{frenchblue}{rgb}{0.0, 0.45, 0.73}
    \definecolor{darkslateblue}{rgb}{0.28, 0.24, 0.55}
    \definecolor{cerulean}{rgb}{0.0, 0.48, 0.65}
    \definecolor{darkpastelgreen}{rgb}{0.01, 0.75, 0.24}

    \newcommand{\cz}[1] { \textcolor{red}{[\hl{cheng:} {#1}}]}
    \newcommand{\hyunc}[1] { \textcolor{burntorange}{[{\hl{hyunc:}} {#1}}]}
    \newcommand{\yax}[1] { \textcolor{magenta}{[{\hl{yaxuan:}} {#1}]}}
    
    \newcommand{\sony}[1] { \textcolor{blue}{[{\hl{songyun:}} {#1}]}}
    \newcommand{\ReviewerFeedback}[1] { \textcolor{brightcerulean}{[{Reviewer Feedback:} {#1}}]}
    \newcommand{\fran}[1]{\textcolor{burntorange}{[{francois:}{#1}}]}
    \newcommand{\rz}[1]{\textcolor{teal}{[{Ruidong: }{#1}]}}
    \newcommand{\lw}[1]{\textcolor{fashionfuchsia}{[{liuwei:}{#1}}]}
    \newcommand{\mose}[1]{\textcolor{burntorange}{[{mose:}{#1}}]}
    \newcommand{\ke}[1] { \textcolor{red!55!yellow}{[{Ke:} {#1}}]}

    \definecolor{CAT-comment}{rgb}{0.95, 0.2, 0.8}
    \definecolor{GL-comment}{rgb}{0.0, 0.54, 0.8}
\newcommand{\theDevice}{SeamPose}

\newcommand{\etal}{et al.~}
\fi

\ifdefined\scifiblind
    \newcommand{\blind}[1]{[omitted for blind review]}
\else
    \newcommand{\blind}[1]{#1} 
\fi

\begin{document}


\title[\theDevice{}]{\theDevice: Repurposing Seams as Capacitive Sensors in a Shirt for Upper-Body Pose Tracking}

\author{Tianhong Catherine Yu}
\email{ty274@cornell.edu}
\affiliation{
  \institution{Cornell University}
  \city{Ithaca}
  \state{New York}
  \country{USA}
}

\author{Manru Mary Zhang}
\email{mz479@cornell.edu}
\authornote{Both authors contributed equally to this research.}
\affiliation{
  \institution{Cornell University}
  \city{Ithaca}
  \state{New York}
  \country{USA}
}

\author{Peter He}
\email{ph475@cornell.edu}
\authornotemark[1]
\affiliation{
  \institution{Cornell University}
  \city{Ithaca}
  \state{New York}
  \country{USA}
}

\author{Chi-Jung Lee}
\email{cl2358@cornell.edu}
\affiliation{
  \institution{Cornell University}
  \city{Ithaca}
  \state{New York}
  \country{USA}
}

\author{Cassidy Cheesman}
\email{crc268@cornell.edu}
\affiliation{
  \institution{Cornell University}
  \city{Ithaca}
  \state{New York}
  \country{USA}
}

\author{Saif Mahmud}
\email{sm2446@cornell.edu}
\affiliation{
  \institution{Cornell University}
  \city{Ithaca}
  \state{New York}
  \country{USA}
}

\author{Ruidong Zhang}
\email{rz379@cornell.edu}
\affiliation{
  \institution{Cornell University}
  \city{Ithaca}
  \state{New York}
  \country{USA}
}

\author{François Guimbretière}
\email{fvg3@cornell.edu}
\affiliation{%
  \institution{Cornell University}
  \city{Ithaca}
  \state{New York}
  \country{USA}
}

\author{Cheng Zhang}
\email{chengzhang@cornell.edu}
\affiliation{%
  \institution{Cornell University}
  \city{Ithaca}
  \state{New York}
  \country{USA}
}

\renewcommand{\shortauthors}{Yu, et al.}

\begin{abstract}
Seams are areas of overlapping fabric formed by stitching two or more pieces of fabric together in the cut-and-sew apparel manufacturing process. 
In~\theDevice{}, we repurposed seams as capacitive sensors in a shirt for continuous upper-body pose estimation.
Compared to previous all-textile motion-capturing garments that place the electrodes on the clothing surface, our solution leverages existing seams inside of a shirt by machine-sewing insulated conductive threads over the seams.
The unique invisibilities and placements of the seams afford the sensing shirt to look and wear similarly as a conventional shirt while providing exciting pose-tracking capabilities. 
To validate this approach, we implemented a proof-of-concept untethered shirt with 8 capacitive sensing seams.
With a 12-participant user study, our customized deep-learning pipeline accurately estimates the relative (to the pelvis) upper-body 3D joint positions with a mean per joint position error (MPJPE) of 6.0 cm.
\theDevice{} represents a step towards unobtrusive integration of smart clothing for everyday pose estimation.
\end{abstract}



\begin{teaserfigure}
  \includegraphics[width=\textwidth]{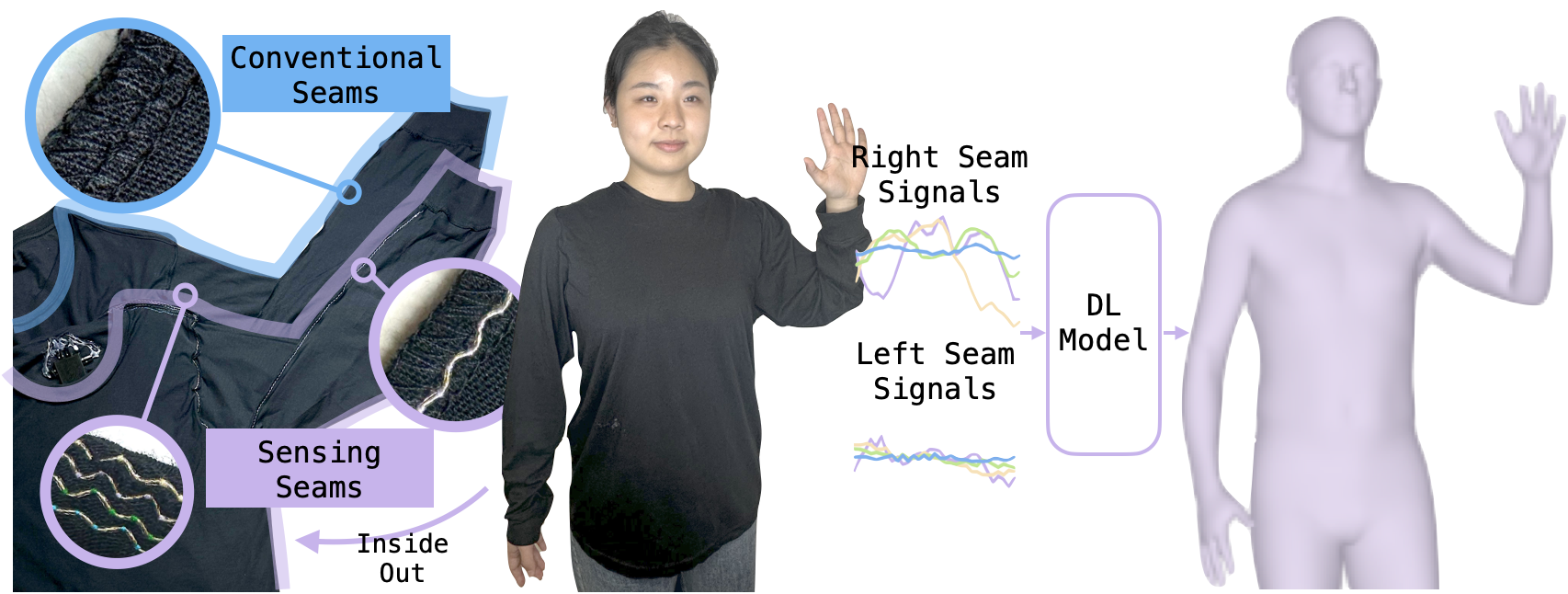}
  \caption{~\theDevice{} repurposes seams as capacitive sensors in a shirt. Without modification to the clothing surface, the sensing shirt looks\&wears similar to a conventional shirt and provides upper-body tracking capabilities. To make the sensing shirt, we machine-sew conductive threads over existing seams. Using 8 channels of capacitive seam signals (4 each on the left/right side) from the shirt, our customized deep-learning model estimates upper-body joint positions.}
  \Description[]{This figure shows the overall system overview of SeamPose. On the left, it shows colorful insulated conductive thread sewn over conventional seams on a black long-sleeve t-shirt. There are 4 different colors used for the thread: purple, blue, green, beige, with each representing a different channel and symmetric sensing channels being the same color. All the seams meet at a device on the upper back portion of the shirt which sends signals wirelessly to a nearby computer for processing. On the right side of the figure, it shows a SeamPose user in a pose along with a glimpse of the signals sent by SeamPose fed into a DL model that creates a 3D upper-body pose estimation of the user.}
  \label{fig:teaser}
\end{teaserfigure}


\maketitle

\section{Introduction}
From the second-to-last Ice Age when humans put on clothes for warmth and protection~\cite{when-clothes-started}, clothes have become indispensable to everyday life.
Not long after Mark Weiser envisioned the future of computers to be woven ``into the fabric of everyday life''~\cite{MarkWeiser}, researchers explored smart clothing for wearable computing~\cite{smart-fabric-1997,TheWearableMotherboard-1999, smart-fabric-1997}.
The always-on nature of clothing makes it an excellent medium for everyday pose tracking, a fundamental task with obvious applications in health care~\cite{clinical-rehab}, human activity recognition~\cite{skeleton-act}, AR/VR interactions~\cite{Oculus}, human-robot interactions~\cite{Manicast}, sports analytics~\cite{pose-golf}, etc.

Clothing deforms, stretches, and shifts as the joints and muscles move.
Prior all-textile wearable movement sensing solutions have exhibited exciting performance by attaching conductive fabric patches across the clothing surface.
More than 10 fabric patches covering different body parts were demonstrated to classify 10 upper-body movements~\cite{undershirt-resistive-gesture} and to track continuous upper-body pose~\cite{Mocapose}.
However, such modifications with patches of conductive fabric alter the base fabric's properties: visual aesthetics and materiality (\textit{e.g.}, softness, stretchability, thickness, and breathability).
As a result, the wearer's experience changes, and the clothing designer needs to be aware of electrode placements' impact on tracking performance~\cite{Mocapose}.

To minimize surface modification and optimize the wearing experience while providing fine-grained tracking capabilities,
we present~\theDevice{}, which repurposes existing seams in a shirt as capacitive sensors for upper-body pose tracking. Seams are areas of overlapping fabric formed by joining two or more pieces of fabric together with stitches. Seam stitching is an essential step in the prevalent and scaled cut-and-sew manufacturing process that produces most everyday apparel. Seam placements are determined by the pattern, a set of templates designed for cutting and sewing the fabric into garment~\cite{more-than-it-seams}. Because seams originally exist on the garment and remain concealed when worn, altering seams with conductive threads will not change the appearance or the materiality of the apparel, while providing exciting tracking capabilities.

While researchers have extensively explored with conductive or functional stitches~\cite{seamless-seams, Skinergy, Embr, Texyz, Sketchstitch, Tessutivo}, as we discussed, large areas of conductive patches are needed for the complex body pose estimation task in the past.
This paper aims to answer the research question: \textbf{whether capacitive sensing seams, repurposed from existing seams, on a shirt can estimate upper-body pose}.

To answer this research question, we developed a proof-of-concept prototype based on a common and basic long-sleeve shirt. To transform the seams into capacitive sensors, we machine-sew insulated conductive threads over the existing seams. It is important to note that we only augment existing seams from the selected shirt pattern and do not strategically add electrodes to locations that better capture body movements~\cite{more-than-it-seams,Mocapaci,StitchedBendsFolds,dancer-motion-resistive}. This prototype is untethered and battery-powered, as shown in Fig.~\ref{fig:PCB}. There are 8 seam electrodes (Fig.~\ref{fig:signals}), 4 each on the left and right side of the shirt. These eight seam electrodes are connected to a customized active capacitive sensing board that measures and transmits the signals wirelessly via Bluetooth. The sensing principle of our approach is that \textit{different body poses and movements will deform seam electrodes and change the coupling between the human body and seam electrodes, leading to unique and complex patterns in measured capacitances.}  

To extract and interpret the pose information from these complex capacitance readings, we customized a deep-learning pipeline that estimates 8 relative (to the pelvis) upper-body joint positions in 3D.  To evaluate~\theDevice{}, we conducted a user study with 12 participants. The results showed that ~\theDevice{} tracks upper body poses with a mean per joint position error (MPJPE) of 6.0 cm, comparable with prior wearable pose tracking systems. In summary, the main contributions of this paper are:

\begin{itemize}
    \item We described a fabrication process to repurpose existing seams in a long-sleeve shirt into capacitive sensors for body pose tracking, one step towards minimally obtrusive conductive textile sensing. 
    \item We developed a proof-of-concept untethered long-sleeve shirt prototype and a deep learning framework that estimates upper body joint 3D positions from the capacitance measured by these conductive seams.
    \item We conducted a user study with 12 participants and achieved promising results, as a proof-of-concept to verify the feasibility of this proposed approach.
    \item We further discussed the opportunities and challenges of generalizing~\theDevice{} on various clothing types and widespread adoption in everyday life.
\end{itemize}

\section{Related Work}

SeamPose tackles pose tracking with capacitive sensing seams in a long-sleeve shirt. Capacitive sensing with textiles exhibited exciting potentials in strain sensing~\cite{batch-cap-sensors-motion}, environmental sensing~\cite{seam-line-cap-multimodal}, object recogniton~\cite{Capacitivo}, gesture recognition~\cite{ProjectJacquard, Pinstripe, Texyz}, etc. Discussing the fundamentals of capacitive sensing is beyond the scope of this paper, we refer readers to Grosse-Puppendahl~\etal and Bian~\etal for comprehensive reviews of body-area capacitive in HCI~\cite{cap-HCI-survey,body-area-capacitive-review}.  In this section, we will focus on discussing the prior work that is closely related to wearable pose-tracking including  1) wearable pose estimation, and 2) body movement sensing with smart clothing.

\subsection{Pose Estimation with Wearable Sensors}\label{sec:rw-pose}
Compared to high-fidelity vision-based MoCap systems~\cite{Vicon,OptiTrack,Humans-in-4D}, wearable solutions enable on-the-go pose estimation without the need for setups in the enviroments.
Xsens~\cite{XSens} attaches 17 wireless inertial measurement units (IMUs) onto a tight-fitting suit to enable professional-grade full-body captures.
Decreasing the number of instrumentation sites alleviates the cumbersome setup but drastically complicates the tracking task due to limited sensed information.
Researchers explored full-body tracking with six IMUs~\cite{DIP, PIP}, six electromagnetic sensors~\cite{Em-pose}, or four flex sensors~\cite{sparseflexsensors} attached onto tight fitting suits.
However, tight-fitting suits are not comfortable for everyday uses.
In AR/VR uses~\cite{poseAR}, full-body poses can be inverse-kinematically inferred with sensors in the headset and hand controllers ~\cite{Avatarposer, Controllerpose, MI-Poser}.
IMUPoser uses IMUs in consumer smartphones, smartwatches, and earphones to infer full-body pose~\cite{Imuposer}.
A single camera could track body pose when mounted on the head~\cite{sceneEgoPose, xr-egopose}/wrist~\cite{BodyTrak} from egocentric/partial body views, respectively.
Recently, smartglasses have utilized ultrasonic sensing to track upper-body poses~\cite{PoseSonic}.
\theDevice{} shares the approach of pose reconstruction with limited sensed information for minimally obtrusive integration and contributes a new approach with seam sensors in clothing that afford everyday uses.

\subsection{Body Movement Sensing with Smart Clothing}\label{sec:textile}
Clothing covers a large area of human bodies and the always-on nature of clothing affords always-on body movement sensing.
Attaching distributed miniaturized electronics (e.g., IMUs~\cite{DIP} and flex sensors~\cite{sparseflexsensors}) to the garments hinders the softness and requires either wireless sensor network~\cite{XSens} or optimized on-body wirings~\cite{wiring-layout-on-suit}.

In contrast, all-textile systems, commonly employing conductive fabrics/threads, augment garments with motion-sensing capabilities while preserving the all-textile softness. 
Liang~\etal bonded conductive fabric patches onto tight-fitting leotards to coarsely monitor dance movements with resistive changes~\cite{dancer-motion-resistive}.
Esfahani~\etal sewed 11 polymerized resistive patches on an undershirt to classify 10 upper body movements ~\cite{undershirt-resistive-gesture}.
Most resistive strain-based approaches require tight-fitting garments, similar to that of IMU suits, because the sensors must be firmly attached to the expected body locations~\cite{onbody-flex-shift}.
Gioberto~\etal used resistive coverstitch on loose-fitting garments to detect fabric bends and folds~\cite{StitchedBendsFolds}.  
MoCapaci integrated 4 textile cables as capacitive antennas into a blazer and effectively classified 20 upper-body poses~\cite{Mocapaci}.
In contrast to works above that analyze the correlation between signals and movements or perform classification tasks, \theDevice{} aims to infer upper-body joints in 3D.

The most recent work, Mocapose~\cite{Mocapose}, is the only existing loose-fitting all-textile system that accomplished the upper-body 3D joint reconstruction task.
With capacitive signals from 16 channels (8 each on the left/right) of conductive fabric patches glued onto a jacket, a 2D CNN-based deep learning model demonstrated strong performance in tracking upper body pose. In contrast, our proposed approach alters the garment at the thread-level instead of the at the fabric-level, a step towards minimizing the alteration of the garment visually and tactilely. By only sewing over the 8 existing seams with conductive threads, \theDevice{} can track the upper body pose continuously on a loose-fit long-sleeve shirt in real-time.


\section{Seams \& Patterns}\label{sec:seam-patterns}
\begin{figure}[h]
    \centering
    \includegraphics[width=\columnwidth]{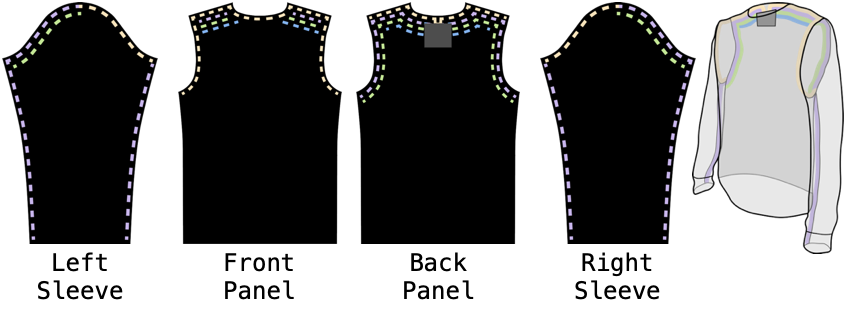}
    \caption{The seams and patterns of our proof-of-concept prototype with a long-sleeve T-shirt. The black fabric pieces represent the patterns of the T-shirt, while the dotted lines are the repurposed seams, also indicating where the stitches are when joining the fabric pieces. The color of the dotted lines can be mapped to the lines, or seams, on the constructed T-shirt.}
    \Description[]{This figure is a simplified view of the four black fabric pieces: left sleeve, front panel, back panel, and right sleeve. These pieces compose the pattern of a long-sleeve black t-shirt used and the figure indicates with colored dotted lines, which match the sensing channel colors, where the conductive thread is sewn on the fabric over the existing seams.}
    \label{fig:pattern}
\end{figure}
For garments, \textit{patterns} serve as templates representing the shapes and sizes of the fabric pieces required to construct a specific garment. \textit{Seams}, on the other hand, are the areas where the fabric pieces overlap when sewn together to make the garment (Fig.~\ref{fig:pattern}). Consequently, patterns define the number of seams, seam placements, and the shapes of fabric to be seamed.
Patterns affect the garments's structure, size, fit, etc.
Although seams vary in their placements, quantity, and length, they are generally distributed throughout the garment. This distribution provides opportunities to collect rich information regarding the body. As a result, in this paper, we leverage this feature and repurpose the seams as capacitive sensors to track body poses.

Because we are repurposing existing seams as sensors, patterns further decide the sensors':
\begin{itemize}
    \item Placements: jackets, blazers, and collard shirts commonly have seams in the front and back of the torso for fit, but T-shirts, sweaters, and sweatshirts generally do not;
    \item Quantity: jackets could easily have twice as many seams as T-shirts, and even for a long-sleeve T-shirt, some patterns have 8 seams\footnote{https://a.co/d/6mg3bW8} while others\footnote{https://www.uniqlo.com/us/en/products/E460354-000} have 10 seams (2 additional ones on the sides of the torso).
    \item Lengths: the sleeve seam in a long-sleeve shirt covers the elbow, an important joint to track, while a short-sleeve shirt does not.
\end{itemize}

For the proof-of-concept prototype, we chose a long-sleeve T-shirt\footnote{https://www.michaels.com/product/long-sleeve-crew-neck-adult-t-shirt-by-make-market-M20033775}.
Long-sleeve T-shirts are basic and common.
They have the fewest numbers of seams among cut-and-sew long-sleeve tops.
Our selected shirt has 5 seams on each side (Fig.~\ref{fig:pattern}): 1 above the shoulder, about 11cm long, 1 in front of the shoulder, about 30cm long, 1 behind the shoulder, about 30cm long, 1 along the sleeve, about 49cm long, and one on the side of the torso, about 44cm long.
To ensure minimal alteration and maximal generalizability, we chose not to repurpose the ones on the sides of the torso because not all long-sleeve shirt patterns have those seams, as explained above.

Although our trained models do not aim to generalize across different patterns, adding more seam electrodes will provide additional information that likely will further improve the tracking performance. The purpose of this paper is to demonstrate the feasibility of this approach and present a baseline for future exploration.  In addition, we acknowledge that the results presented in our paper can not be directly replicated on sleeveless garments like strapped and strapless tops.

\section{\theDevice{} Implementation}\label{sec:impl}
We repurpose seams as capacitive sensors for upper-body pose tracking, without modifying the clothing surface.
\theDevice{} prototype has three main components: 
\begin{itemize}
    \item A shirt with 8 conductive seam electrodes (detailed in Sec.~\ref{sec:impl-fab}): the electrodes are symmetric on the left/right side of the shirt with 1 above the shoulder, 1 in front of the shoulder, 1 behind the shoulder, and 1 along the sleeve;
    \item A customized capacitive sensing board, detailed in Sec.~\ref{sec:impl-hardware}; and
    \item A deep learning pipeline estimating the positions of 8 upper body joints from the readings of capacitive sensors (Sec.~\ref{sec:impl-signals}), detailed in Sec.~\ref{sec:impl-dl}.
\end{itemize}

\subsection{Conductive Seams Fabrication}\label{sec:impl-fab}
\begin{figure}[t]
    \centering
    \includegraphics[width=\columnwidth]{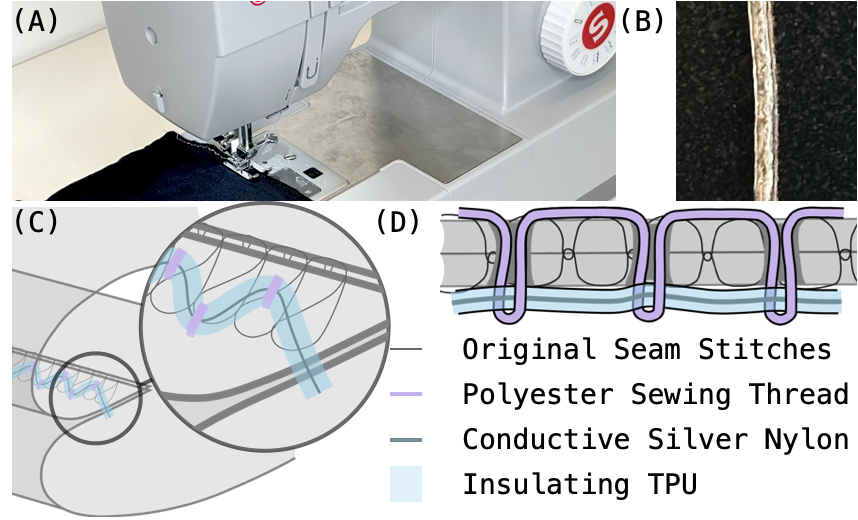}
    \caption{Machine Sewing Conductive Thread. (A) Home sewing sewing machine setup. (B) The off-the-shelf insulated conductive thread in use. (C) An illustration of machine-sewn conductive thread traces over existing seams, overlapping areas formed when stitching (by the original seam stitches) pieces of fabric together. (D) A side-view illustration of how the TPU-insulated conductive thread with silver nylon core is stitched onto the fabric.}
    \Description[]{This figure is composed of 4 subfigures: A, B, C, and D. A shows the home sewing machine setup. B is an image of the off-the-shelf insulated conductive thread used by SeamPose. C shows an illustration of how the machine-sewn conductive thread is traced over existing seams on the t-shirt. D shows a side-view illustration of how the TPU-insulated conductive thread with a silver nylon core is stitched onto the fabric. In both C and D the insulating TPU is highlighted in blue, the polyester sewing thread is purple, and the conductive silver nylon is dark blue.}
    \label{fig:fab}
\end{figure}

To transform conventional seams into capacitive sensing seams, we machine-sew conductive thread over existing seams, as shown in Fig.~\ref{fig:fab}.
The base fabric of our selected unisex shirt (Make Market Long Sleeve Crew Neck Adult T-Shirt) of size medium is 100\% single knit cotton jersey.
We use a home sewing machine (SINGER Heavy Duty 4423 Sewing Machine) with a sewing needle of size 80/12 (SCHMETZ Universal 130/705).
The top thread uses a conventional polyester sewing thread.
Similar to prior works that use functional threads/wires as bobbin threads to relieve mechanical stress~\cite{Texyz,Skinergy,seamless-seams}, the bottom bobbin thread uses an off-the-shelf TPU-coated 2-ply silver-plated nylon thread (Shieldex 117/17 x2 HCB TPU, Fig.~\ref{fig:fab}(B)) with a resistance profile of  $< 300 \Omega$/m.
We choose a zigzag stitch, common for seaming with home sewing machines.
After extensive experiments for a clean and consistent finish, we set the thread tension, stitch length, and stitch width to 5, 2.5, and 2.5, respectively.
\paragraph{Conductive Thread Selection}
The choice of conductive thread can heavily impact the overall system performance.
In our early prototypes, we experimented with a non-insulated silver-plied nylon thread (LessEMF, <100 Ohm/cm).
We noticed (a) electrode shorting caused by body movements so we looked into insulated conductive threads that are compatible with home sewing machines, and (b) frequent conductive thread breakage when sewing.
For our selected TPU-coated thread, the elastic TPU coating (a) insulates the conductive core, and (b) provides strong mechanical properties allowing $21\%\pm5$ elongation before breaking\footnote{https://www.shieldex.de/wp-content/uploads/2021/05/Y-VTT-Datasheet-Shieldex-117-17-x2-HCB-TPU-V4.pdf}.
When prototyping with the TPU-coated thread, the conductive thread never broke when sewing.
The strong mechanical property not only increases the fabrication success rate but also strengthens the prototype's durability, important for wearable applications with frequent rewearing. 
Our final prototype has been tested with over 20 people and hundreds of remounting without breaking.
\begin{figure}[t]
    \centering
    \includegraphics[width=\columnwidth]{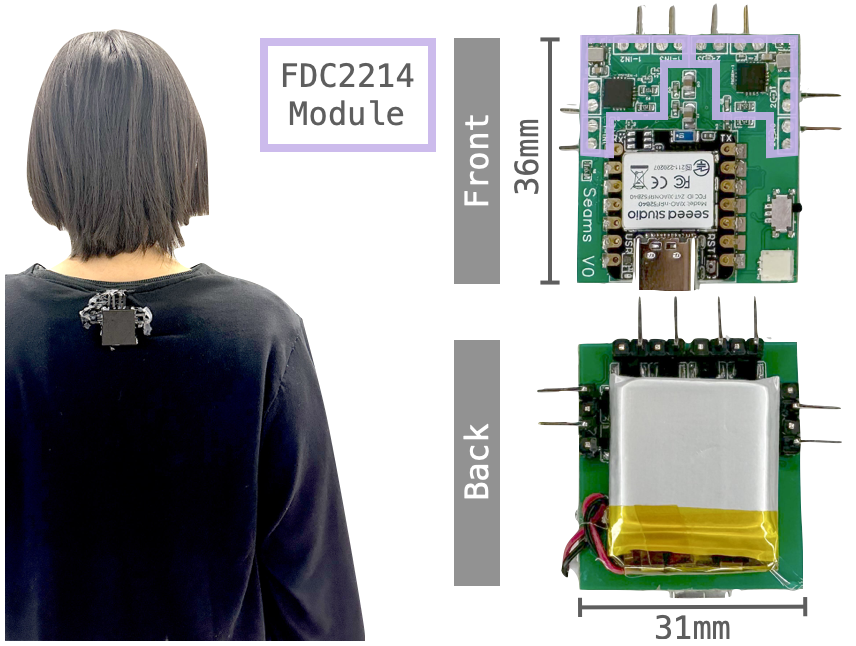}
    \caption{The battery-powered customized sensing board is housed inside a 3D-printed PLA case and hot-glued onto the prototype below the neck.}
    \Description[]{This figure has two parts, on the left side it shows a picture of the back of a user wearing SeamPose. From the outside, SeamPose looks like a normal black long-sleeve t-shirt, except for the 3D-printed customized sensing board housing hot-glued below the neck. On the right side of the figure, there are two images of the customized sensing board. The top image shows the front of the sensing board and a length measurement of 36mm. The bottom image shows the back of the sensing board and a width measurement of 31mm.}
    \label{fig:PCB}
\end{figure}
\begin{figure*}
\includegraphics[width=\textwidth]{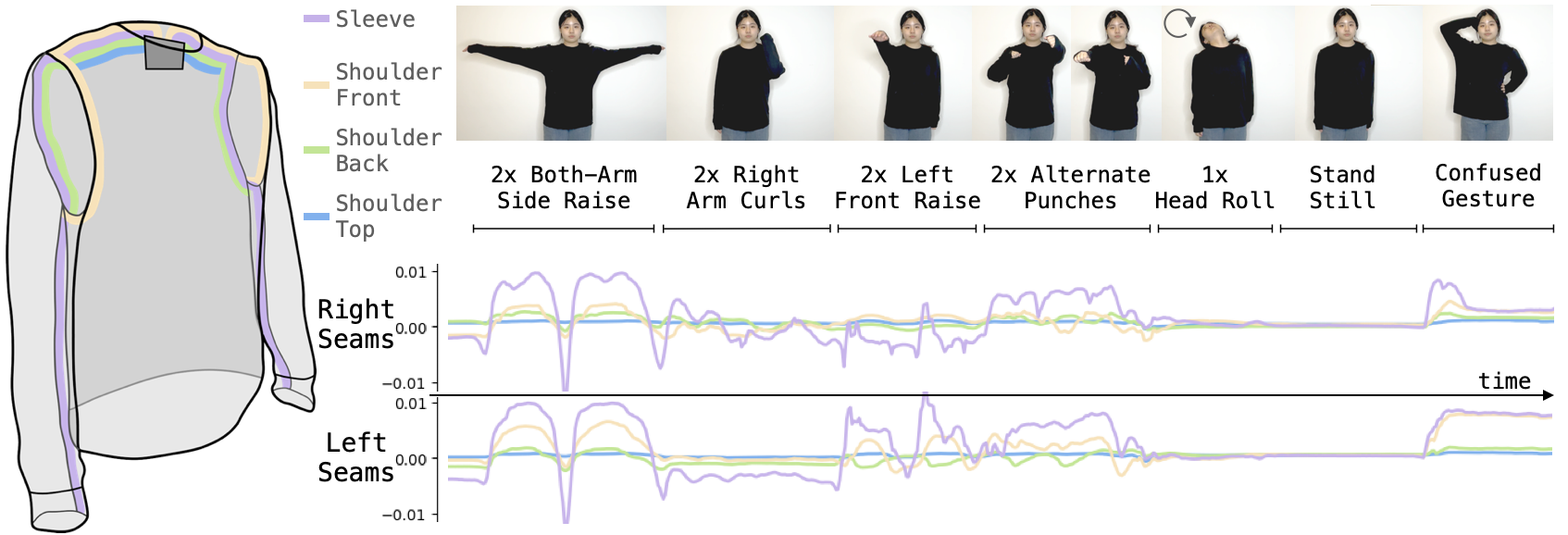}
  \caption{Example~\theDevice{} Signals. We show 19 seconds of continuous~\theDevice{} the left and right seam signals (4 each) as the wearer performs symmetric (both arm raises) arm movements, asymmetric arm movements (right arm curls, left front raises, alternate punches, confused gesture), head movements, and standing still. The colors map the signals and their corresponding electrode placements, illustrated on the left. The numerical value of signal on $y$-axis is median normalized.}
  \Description[]{On the left side of this figure, it shows a drawing of SeamPose with the sensing electrodes highlighted and color-coded: Sleeve is purple, Shoulder Front is beige, Shoulder Back is green, Shoulder Top is blue. On the top right side, there are images of a user on top performing various movements with graphs of the Left and Right seams at the bottom displaying the corresponding SeamPose signals. The colors of the signals correlate with the sensing channels.}
  \label{fig:signals}
\end{figure*}
\paragraph{Connectors}
Electrically and mechanically robust connections between textile-based sensors and rigid electronics remain an open research problem~\cite{textile-connector-review}.
We connect our insulated threads similar to insulated wires with DuPont wire-to-wire connectors.
We crimp the stripped conductive thread core inside the conductor tab and the insulating TPU-coated thread inside the insulation tab. This method ensures stable connections during movements and re-wearings.

\subsection{Customized Sensing Board}\label{sec:impl-hardware}

The purpose of the customized sensing board (Fig.~\ref{fig:PCB}, 36x31mm) is to measure and transmit the capacitance of the connected conductive seams without hindering body movements.
We replicate the single-ended configuration circuit in FDC2214 (Texas Instruments) evaluation module\footnote{https://www.ti.com/tool/FDC2214EVM}: the 4-channel capacitance-to-digital converter measures capacitances with great resistance to electromagnetic interference.
We chain 2 FDC2214s on I2C to read 8 channels, at a configured sample rate of 32Hz.
XIAO nRF52840's (Seeed Studio) transmits the sensor readings via onboard Bluetooth Low Energy (BLE) functions to a nearby computer.
We use Arduino to program the firmware and UART for Bluetooth communication.
We power the circuit with a 3.7V 290mAh Lipo battery.
The mean measured\footnote{https://lowpowerlab.com/guide/currentranger/} power consumption is 12.7mA, which can continuously transmit measurements for 22.83h.
Approximately, the board weighs 11g and costs US\$27.

\subsection{\theDevice{} Signals}\label{sec:impl-signals}
Changes in the wearer's upper body pose cause changes in the seam electrode's sensed capacitances, which are input into our deep learning pipeline that infers the relationship between the body pose and measurement capacitances.
When connected to the sensing board, the conductive seams become self-capacitance sensor electrodes.
In a loading-mode capacitive sensing system, the setup is simple: each sensor electrode acts as both a transmitter and a receiver~\cite{first-loading-mode}.
In any loading-mode system, electrical current displacement occurs as the electrode's capacitive coupling with the surrounding changes. 
In~\theDevice{}, the seam electrodes are coupled with the wearer's body, so when the body pose changes (\textit{i.e.}, limb movements), the seams move along with the limbs, leading to changes in the coupling between the body and seam electrodes, hence the current displacement changes.
In addition, the seam electrode itself deforms, distorts, and displaces during body movements, altering the electrode's capacitance which also affects the current displacement. 
In a nutshell, we conjecture the capacitive changes are attributed to (1) coupling changes between the seam electrodes and the wearer's body, and (2) seam electrode's self-capacitance changes.

Notice that in Fig.~\ref{fig:signals}, the sleeve electrodes (purple) not only trace along the sleeve seams but also trace along the seams in the front (beige) of and on top of the shoulder (blue).
This is a design choice to avoid additional wire routings across the shirt: one ends of all electrodes meet at the sensing board.
Even though some sensor placements "overlap", they still individually provide sensing information, which we further validate in Sec.~\ref{sec:ablation}.

As shown in Fig.~\ref{fig:signals}, the seam signals correlate to the wearer's movements.
The sleeve electrodes, colored purple, have the largest changes in magnitude as they are the longest electrodes with the largest range of motion.
The shoulder front (colored beige) and back (colored green) electrodes have similar magnitudes as they are symmetrically placed, but the relationship between these two signals depends on the motion.
For example, for the both-arm side raises, the two signals both increase as the arms raise and decrease as the arms return, but for alternate punching movements, one signal increases as the other decreases.
The shoulder top (colored blue) has the smallest magnitude for its shortest length and relatively stationary nature, compared with other electrodes that touch moving joints.
Though the magnitude is small, it is proven to still be information-rich in our later evaluation.
\begin{figure*}
  \includegraphics[width=\textwidth]{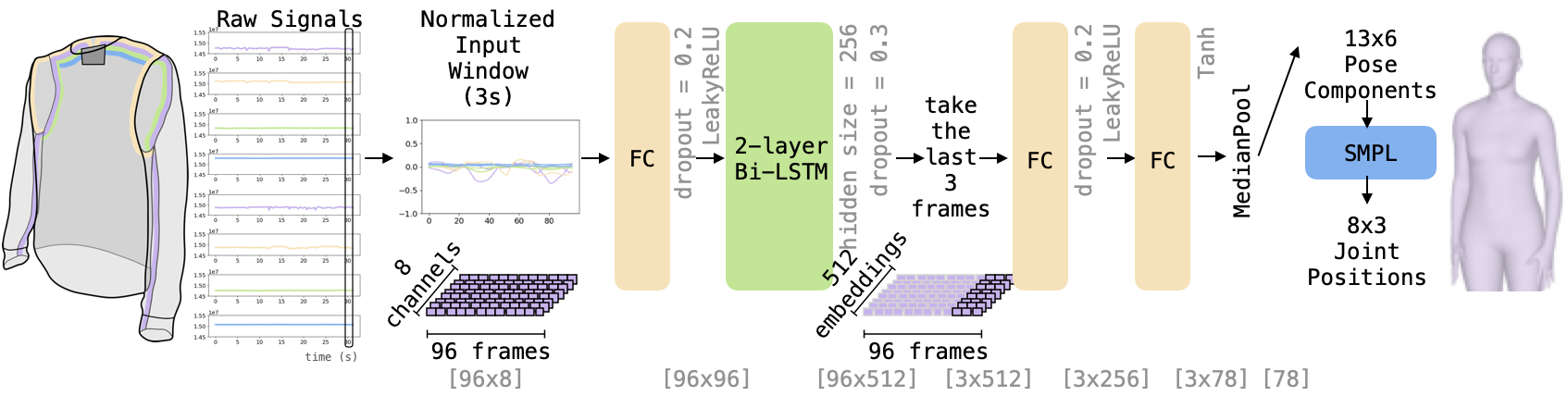}
  \caption{Customized Deep Learning Pipeline.}
  \Description[]{The figure shows the Customized Deep Learning Pipeline used by SeamPose. The raw 8-channel signals are put into a normalized input window with 96 frames, equivalent to 3 seconds. This is then fed into an FC layer with a dropout of 0.2, activated by LeakyReLU. Then the embeddings go through a 2-layer Bi-LSTM with a dropout of 0.3 and hidden size of 256. The last 3 frames, out of 96 frames, of the output 512 embeddings are then fed into a fully connected later layer with a dropout of 0.2, LeakyReLU activated. Then the embeddings go through a final fully-connected layer, activated by Tahn. The output of dimension 3x78 goes through medianPool into dimension 78 and corresponds to 13x6 pose components. Finally based on the 13x6 pose components, SMPL calculates 8x3 joint positions.}
  \label{fig:DL}
\end{figure*}

For asymmetric movements (\textit{e.g.}, right arms curls and left front raises), the static side still exhibits signal changes with small magnitudes. 
Moving one side of the shirt often stretches the other side because different parts of the shirt are interconnected and body movements involve complex muscle engagements.
For example, a one-arm raise could (a) stretch the back of the shirt which changes the signals as the sensing board is in the back, and (b) alter the whole body’s coupling with the shirt.
Further, at the thread-to-sensing-board connection, some insulated threads are close to and even overlap with each other and thus attributed to crosstalks, which is a limitation to be addressed in the future, but in our prototype, we input the 8 seam signals together, instead of side-by-side, to avoid possible confusion to the DL model.

Although our prototype does not cover or directly instrument the head, it can sense the head movement (see head roll signals differing from standing still signals in Fig.~\ref{fig:signals}) based on overall body-capacitance changes and subtle shoulder movements. 
To extract and interpret the pose information from these complex capacitance changes, we developed a data-driven approach.

\subsection{Deep Learning Pipeline}~\label{sec:impl-dl}
With eight channels of capacitive signals from the seams on the shirt, we estimate the wearer's upper body pose in 3D, relative to the pelvis.
In the previous subsection, we show the correlation between body movements and seam signals, but the complex mapping between the 8 channels and the body pose is not immediately clear. 
The reconstruction task is challenging because the tracked body parts have a total of 19 degrees of freedom~\cite{bodyDOf}, and we only have sparsed 8 channels of 1-dimensional temporal signals, constrained by our design choice that minimally alters everyday clothing.
Unlike tight-fitting suits, the sensors are not firmly fixed to a mapped body part, complicating the problem~\cite{onbody-flex-shift}.
The task is further complicated by the soft cotton fabric having natural draping variations.
For example, when the wearers raise their arms over the head, the sleeves naturally slide down the arms and attribute signal variations.
Embracing these technical hurdles in pursuit of comfortable and seamless pose-tracking integration into everyday clothing, we customized a deep-learning pipeline to learn the complex mapping between the 8 input channels.

\subsubsection{Ground Truth Acquisition}

Recent computer vision advancements in pose reconstruction with a single RGB image have proven more accurate for users in loose-fitting cloth~\cite{GTforLooseWearables}.
Similar to prior works on non-vision-based pose tracking~\cite{CAvatar,Mocapose,li2024eyeecho,eario,lee2024echowrist,yu2024ring}, we selected state-of-the-art computer vision models for ground truth acquisition.
Given an RGB image frame, Detectron2\footnote{https://github.com/facebookresearch/detectron2} first detects the area containing the person, then HMR2.0~\cite{Humans-in-4D}, a vision transformer-based model, estimates the SMPL~\cite{SMPL} representation of the detected person.
SMPL is a mathematical model that describes a human body mesh template of 6890 vertices with shape components $\beta$ (dim($\beta$)=10) and pose components $\theta$ (dim($\theta$)= 216 = 24 joints x (3 x 3) rotation matrix representing the rotation from its kinematic prior).
Among the 24 joint components, the pelvis provides global orientation (\textit{i.e.}, the direction the body is facing) and the other 23 joint components represent rotations from their kinematic prior.
SMPL also acts as a forward-kinematic model and calculates joint positions, $J$, based on joint rotations, $\theta$.
Note, \theDevice{} does not predict body shape or global orientation.
The ground truth labels have 2 folds:
\begin{itemize}
    \item 13 pose components of size [13 x 6]: rotations of 13 upper body joints, and we exclude the hands, similar to~\cite{Avatarposer,MI-Poser}. SMPL uses 3 x 3 rotation matrices to represent rotations, but a 6D rotation representation has been proven effective~\cite{Imuposer,Avatarposer,6D-rotation}, so we convert rotation matrices into 6D rotation vectors.
    \item 8 joint positions of size [8 x 3]: 3D position representation of upper-body joints, in x,y,z coordinates, calculated by SMPL with pose components: nose, neck, right shoulder, right elbow, right wrist, left shoulder, left elbow, and left wrist. We scale the joint positions into physical units (meters) with measured arm lengths and center the pelvis at the origin.
\end{itemize}

\subsubsection{Input Normalization}
The input to the model is a 3s-window (96 frames) of 8 channels of seam signals.
The capacitive readings depend on the body pose and the capacitance of the seam electrodes.
Because the electrodes are not of equal length and are distributed on different parts of clothing, channels of signals are not centered together.
We first calculate the median of the most recent window of 5.6s (180 frames) and perform a median normalization.
And to better center individual channels around 0, we subtract each channel with: 0.98 x channel median within the 3s-window.
The scaling factor aims to preserve inter-channel relationships.
For example, without the scaling factor, standing still for more than 5.6s with arms down and arms up will have the same normalized signals.

\subsubsection{Model}
The model architecture is detailed in Fig.~\ref{fig:DL}.
The input signals, of dimension [96x8], are first transformed into an embedding of dimension 96 with a linear layer.
Then, the embeddings go through a 2-layer bidirectional LSTM (inspired by~\cite{Imuposer,sparseflexsensors,DIP,LIP}) of hidden dimension 256.
LSTM is chosen to better model the temporal patterns, but there are potential optimizations on model architectures (e.g., transformers) for future works.
We take the last 3 prediction frames (0.09s) and decode the embeddings with 2 linear layers with out-feature sizes of 256 and 78 (for the 13 pose components).
We then \texttt{MedianPool} the last 3 prediction frames to mitigate jitter~\cite{sparseflexsensors}.
Finally, we use SMPL to calculate joint positions based on the pose components.
Our model directly regresses the pose components representing joint angles that are more transferrable across body sizes.
The conversion to joint positions allows easy comparisons with other systems as the position error is the common metric for wearable pose tracking.
We further smooth the joint position predictions with a running median filter of window size 5 (0.17s).

\subsubsection{Training}\label{sec:training-scheme}
We implemented the models in PyTorch and trained them on an NVIDIA GeForce RTX 2080 Ti.
We use the Adam Optimizer with a cosine learning rate scheduler.
To compute the loss, we add the mean absolute error (MAE) losses of pose components and joint positions:
\begin{equation}\label{eqn:Voc}
    \mathcal{L} = ||\theta - \theta^{*}||_{1} + ||J - J^{*}||_{1}
\end{equation}
Here, $\theta$ and $J$ are the predicted 13 pose components and 8 joint positions.
$\theta^{*}$ and $J^{*}$ are the ground truth 13 pose components and 8 joint positions.
The batch sizes are 512.

We adopt a two-stage training scheme: (1) User-Independent training stage: we first train a user-independent model with 15 epochs and a starting learning rate of 8e$^{-3}$; and (2) User-Adaptive training stage: we then finetune the trained user-independent model with another 10 epochs and a starting learning rate of 4e$^{-4}$.

\subsubsection{Data Augmentation}
To improve the model's robustness against different body sizes and movement patterns, we apply data augmentation techniques to introduce variations to the training data in each epoch.
To mitigate window and channel median variations that impact the input normalization, at 80\% chance, we apply random shifts to window and channel medians independently.
To account for signal range variations caused by the fit of the shirt, at 80\% chance, we apply random scaling in the range of [94\%, 106\%] to the normalized input.
To further introduce randomness, at 80\% chance, we scale individual reading in the range of [99.7\%, 100.3\%].

\subsection{Real-Time Inference Pipeline}
\begin{figure}[t]
    \centering
    \includegraphics[width=\columnwidth]{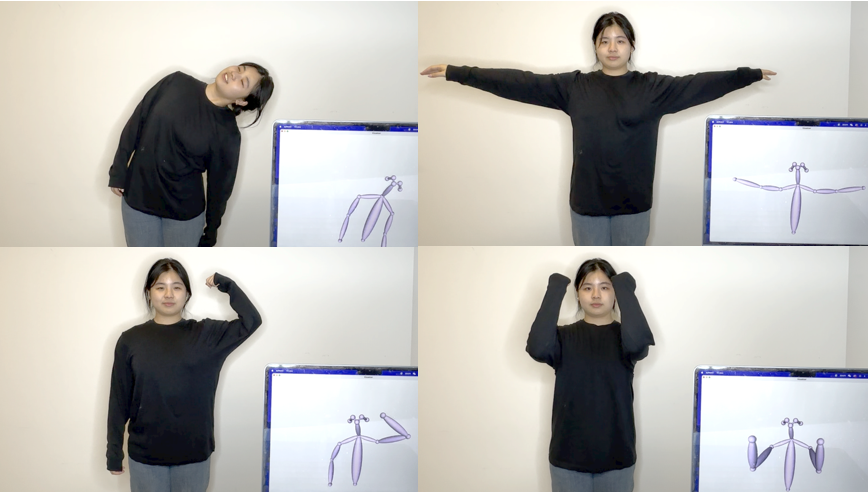}
    \caption{Real-Time Inference with \theDevice{}. Note, for our MPJPE calculation, the head is represented by only the nose key point to avoid error dilution, but for visualization purposes, we include eyes and ears key points reconstructed by SMPL.}
    \Description[]{This figure shows four example screenshots of real-Time inference with SeamPose: with the wearer wearing SeamPose on the left and predictions on a laptop screen on the right. In the first example, the user is learning to their left. In the second example, the user is T-posing. In the third example, the user is doing a bicep curl with their left arm. In the fourth example, the user is in a boxing stance.}
    \label{fig:live-inference}
\end{figure}

We implement a real-time end-to-end inference pipeline (Fig~\ref{fig:live-inference}).
For real-time joint visualization, we use the visualizer provided by EasyMocap~\cite{easyMocap}.
On an Apple Macbook Air (2022), the inference (10.7ms) and visualization (16.0ms) latency are 28ms in total. 

\section{User Study Evaluation}\label{sec:user-study}
To evaluate ~\theDevice{}'s continuous upper-body pose estimation performance, we conducted a user study, approved by the Institutional Review Board (IRB).
We recruited 12 participants (6 self-identified as male, 6 as female, mean age=24.9, std age=4.0) spanning a variety of body shapes, detailed in Table~\ref{tab:measurments}. 
Each study lasted about 1 hour and compensated US\$15.

We conducted the study in an experiment room on a university campus. At the beginning of the study, the experimenter instructed the participant to stand in front of a green screen. A laptop (Apple Macbook Air, 2022) was placed on a desk about 3m away from the participant.
The laptop continuously recorded (a) ground truth video via its built-in camera (30fps) and (b) ~\theDevice{} sensing data received via Bluetooth from the prototype described in Sec.~\ref{sec:impl} and synchronized the two with its clock.
The laptop and its connected monitor displayed visual stimulus for movements and the camera view that captures the full body. We chose video stimuli over photos and text as videos contain details of the movements. Participants were asked to follow the movements on the video but were not strictly asked to follow the same pace nor the exact movement patterns (\textit{e.g.}, golf\&tennis swings and dance movements varied greatly among participants). 

For each participant, we collected 8 sessions of data.
Before each session, the participant was instructed to take off and put back on the shirt themself in order to evaluate our system across different wearing sessions. Participants wore their own clothes of various types and sizes that fit underneath our prototype shirt. Each session lasted about 227 seconds.
Each session contains 3 sections, including three types of upper-body movements informed by prior work: 
\begin{figure*}
\includegraphics[width=\textwidth]{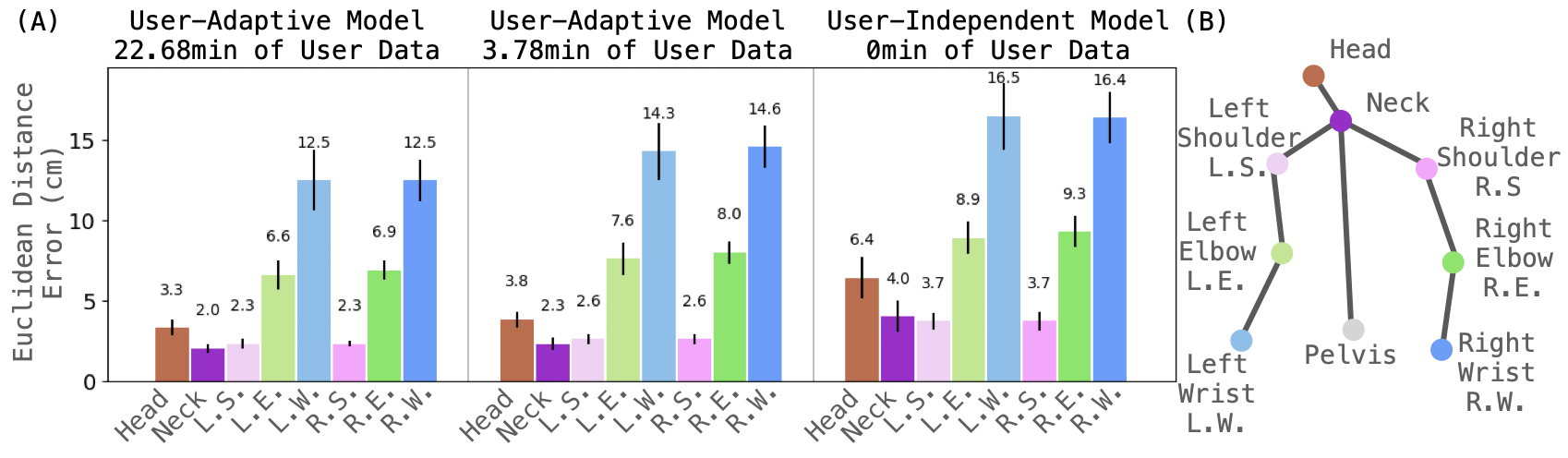}
  \caption{Joint Error Breakdown. In (A), we show the distance error distributions of the 8 predicted joints, labeled in (B). Error bars represent the standard deviation across participants.}
  \Description[]{This is composed of 2 subfigures: A and B. A shows 3 color-coded bar graphs with the distance error distributions of: a User-Adaptive Model with 22.68 minutes of User Data, a User-Adaptive Model with 3.78 minutes of User Data, and a User-Independent Model with 0 minutes of User Data. The Y-axis of each graph is the Euclidean Distance Error in centimeters. The X-axis of each graph has the data points for: Head, Neck, Left Shoulder, Left Elbow, Left Wrist, Right Shoulder, Right Elbow, Right Wrist. For the user adaptive model with 22.7 min of training data, the user adaptive model with 3.78 min of training data, and the user-independent model with 0 min of user data, respectively, the head errors are 3.3, 3.8, and 6.4 cm; the neck errors are 2.0, 2.3, and 4.0 cm; the left shoulder errors are 2.3, 2.6, and 3.7cm; the left elbow errors are 6.6, 7.6, and 8.9cm; the left wrist errors are 12.6, 14.3, and 16.5cm; the right shoulder errors are 2.3, 2.6, and 3.7cm, the right elbow errors are 6.9, 8.0, and 9.3cm; the right wrist errors are 12.5, 14.6, and 16.4cm. B shows a stick-figure representation of a user’s upper body with each of the joints labeled and color-coded, corresponding to the labels in (A).}
  \label{fig:joints-results}
\end{figure*}
\begin{itemize}
    \item  \textbf{Section 1: 54 randomized unique movement videos (195s)}, detailed in Appendix~\ref{sup:video-descriptions}, cover casual/daily gestures~\cite{Mocapaci}, sports movements~\cite{MI-Poser, sparseflexsensors}, and controlled terminal poses~\cite{PoseSonic} movement sequences~\cite{Mocapose} which explore poses that are uncommon in daily activities but kinematically feasible.
    \item \textbf{Section 2: 2 TikTok dance videos\footnote{https://www.tiktok.com/@sophielaverie/video/7334053266050911521}$^{,}$\footnote{https://www.tiktok.com/@sophielaverie/video/7326633661162474784} (21s)} introduce fluent movements and rare poses~\cite{Mocapose}.
    \item \textbf{Section 3: Freestyle movements (10s)} introduce unseen poses/movements that are not included in our defined set. Participants are instructed to perform random movements of their choice, not limited to the upper body.
\end{itemize}
As a proof-of-concept on a research prototype, we can not exhaust all possible upper-body poses. The three different types of movements are chosen to provide enough variance of the body poses to demonstrate the potential of this proposed sensing approach to tracking body poses. 

In total, we collected 1813.6 seconds of data from each participant, resulting in $1813.6\text{s} \times 30 \text{fps (camera's sampling rate)} = 54408$ training/testing instances. From all 12 participants combined, we collected 6.03 hours of data.

At the end of the study, participants completed a questionnaire collecting information about their demographic, body sizes (measured by the experimenter), and the prototype's wearability.

\section{Results}\label{sec:results}
We evaluated \theDevice{} for both user-adaptive and -independent scenarios, as well as estimating performance for each body joint and various motions. Additionally, we conducted an ablation study to understand the impact of seam placements on \theDevice{}'s continuous tracking performance.

\paragraph{Evaluation Metrics}
Informed by other wearable pose tracking systems~\cite{BodyTrak,PoseSonic,Mocapose,MI-Poser}, we adopt Mean Per Joint Position Error (MPJPE) as the evaluation metric for continuous pose tracking: the mean Euclidean distance errors of 8 joint positions in centimeters (cm), relative to the pelvis.

\begin{figure}[t]
    \centering
    \includegraphics[width=\columnwidth]{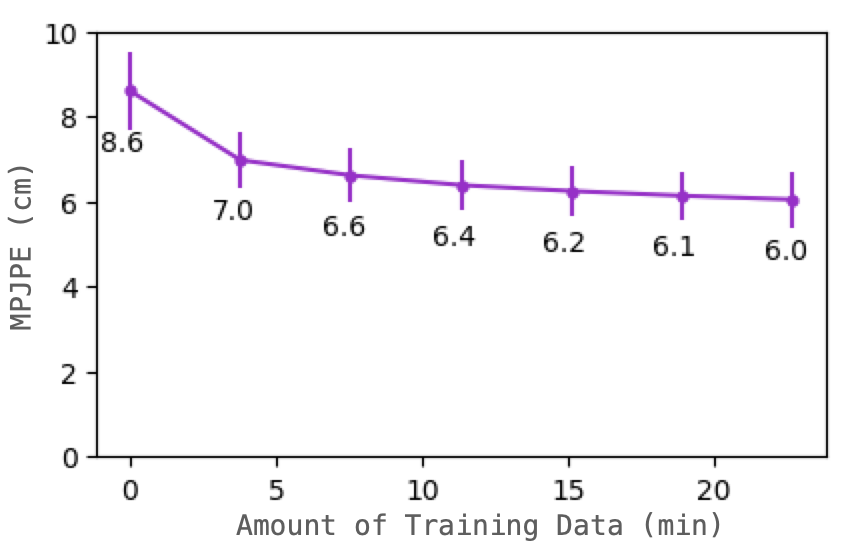}
    \caption{As the user provides more training data for the user-adaptive model, the model's tracking performance increases. Error bars represent the standard deviation across participants.}
    \Description[]{This figure shows a line graph that shows how the model’s tracking performance increases when more training data. The y-axis is the Mean Per Joint Position Error in centimeters and the x-axis is the amount of training data in minutes at a increment of 3.7min: the error is 8.6cm 0 minutes of training data; the error is 7.0 cm with 3.7min of training data; the error is 6.5cm with 7.4 min of training; the error is 6.4cm with 11.1min of training data; the error is 6.2 with 14.8min of training data; the error is 6.1cm with 18.5min of training data; the error is 6.0 with 22.7min of training data.}
    \label{fig:ft-curve}
\end{figure}

\subsection{User-Adaptive Model Results}
As mentioned in Sec.~\ref{sec:training-scheme}, we first trained a user-independent model, without the participant's data in the training set, and then we fine-tuned the user-adaptive (UA) model with the the evaluated participant's data.
To simulate the user calibrating the device before they start using it, we use the first 6 sessions (22.7 min) of data from the evaluated participant for fine-tuning and testing on the last 2 sessions.
Note that the participant took off and put back on the shirt before every session, our system is session-independent.
~\theDevice{} achieves an overall MPJPE of 6.0cm (std=0.65cm).
We recognize that providing 22.7 min of calibration data may not be always preferred for the optimal user experience. Therefore, we conducted further experiments exploring how much fine-tuning data is required to achieve a good tracking performance.
As shown in Fig.~\ref{fig:ft-curve}, reducing the fine-tuning data to 18.9min, 15.1min, 11.3min, 7.6min, and 3.8min, MPJPE increases to 6.1cm, 6.2cm, 6.4cm, 6.6cm, and 7.0cm, respectively.
With only 3.8 minutes of user-specific training data that contains only 1 repetition of each movement,~\theDevice{} still has a promising tracking performance of 7.0cm (std=0.68cm), 1cm worse than that with 22.7min of training data.

\subsection{Unseen User: User-Independent Model Results}\label{sec:unseen-user}
For an even better user experience, the shirt should be able to track body poses "out-of-box" without any calibration data from the new user, \textit{i.e.}, user-specific training data.
We adopt a leave-one-participant (LOPO) cross-validation to evaluate our performance in this scenario. User-independent (UI) models were trained 7.75 hours of data: 30.2 minutes of data from each of the other 11 participants and 5 hours of data from 7 researchers.
As shown in Fig.~\ref{fig:ft-curve}, the MPJPE increases to 8.6cm (std=0.93cm), 1.6cm worse than 3.8min of training data, and 2.6cm worse than 22.7min of training data.
The performance degradation is expected because individuals' body sizes and capacitances vary, but~\theDevice{} still yields low-fidelity tracking, sufficient for certain applications like activity recognition and lifelogging.

\subsection{Results Analysis based on Joints}
To help better understand our performance, we further analyzed our performance based on individual joints, as illustrated in Fig.~\ref{fig:joints-results}. 
Similar to prior work~\cite{Mocapose,Imuposer,MI-Poser}, the wrists with the most moving distance have the largest errors: 12.2cm, 14.4cm, and 16.4cm with 22.7min, 3.8min, and 0min of user data respectively.
The end effectors accumulate errors along their long kinematic chains and the wrists have large ranges of movements.
The elbows have the second largest errors: 6.7cm, 7.8cm, and 9.1 cm with 22.7min, 3.8min, and 0min of user data. The neck and shoulders have the smallest errors, as they did not have much movement compared to wrists and elbows. 
We notice a big difference in estimating the position of the head between the UI model (6.4cm) and UA model (3.8cm) with 3.78min of data. We conjecture this is caused by the fact that ~\theDevice{} does not directly instrument the head, unlike other tracked joints that affect the seam electrode shapes, so head movements are mostly inferred from the overall coupling change between the body and the seams. This indicated that our system may need calibration data for a reliable estimate of head position. 
\begin{figure}[t]
    \centering
    \includegraphics[width=\columnwidth]{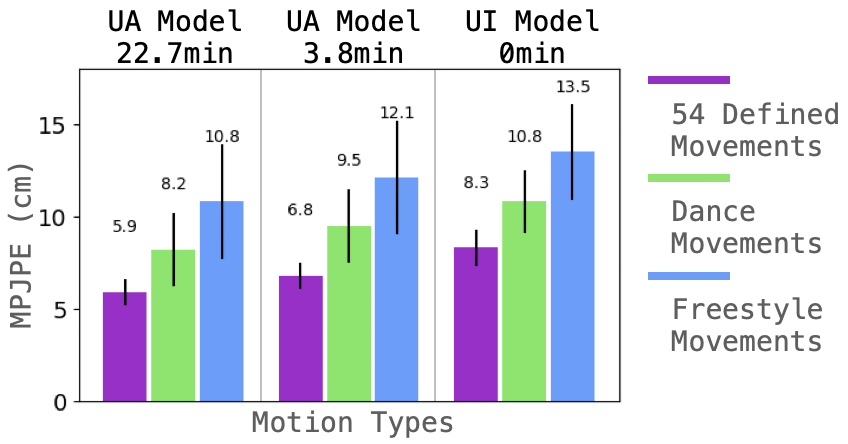}
    \caption{Movement Error Breakdown. Error bars represent the standard deviation across participants.}
    \Description[]{This figure shows 3 bar graphs with movement error break of: a User-Adaptive Model with 22.68 minutes of User Data, a User-Adaptive Model with 3.78 minutes of User Data, and a User-Independent Model with 0 minutes of User Data. The y-axis is the Mean Per Joint Position Error (MPJPE). The bar graph bars are color-coded with purple being the 54 defined movements, green being the dance movements, and blue being the freestyle movements. For the user adaptive model with 22.7 min of training data, the user adaptive model with 3.78 min of training data, and the user-independent model with 0 min of user data, respectively, the defined movements have MPJPEs of 5.9, 6.8, and 8.3 cm; the dance movements have MPJPEs of 8.2, 9.5, and 10.8cm; freestyle movements have MPJPEs of 10.8, 12.1, and 13.5cm.}
    \label{fig:motion-res}
\end{figure}

\subsection{Results Analysis based on Motion}\label{sec:unseen-motino}
We further break down the results into motion types.
Like many other data-driven wearable MoCap systems, the performance decreased for estimating unseen movements. As we discussed in Sec.~\ref{sec:user-study}, we cannot exhaust all possible upper-body movements\&poses for evaluation.
However, when we designed the study, we explicitly included dance and freestyle movements to better understand our system's limitations on extreme and unseen poses and movement patterns.
As shown in Fig.~\ref{fig:motion-res}, across all three models, defined movements (5.9cm, 6.8cm, and 8.3cm with 22.7min, 3.8min, and 0min of user data) have the smallest errors, followed by dance movements (8.2cm, 9.5cm, and 10.8cm with 22.7min, 3.8min, and 0min of user data) and freestyle movements (10.8cm, 12.1cm, and 13.5cm with 22.7min, 3.8min, and 0min of user data). Adding user-specific training data also consistently improves the tracking accuracy for each movement type. This result suggested that the system similar to many data-driven systems, can further benefit from being trained with a much larger dataset containing a large variety of body movements. However, as a proof of concept, it still showed reliable tracking performance on unseen movements and subjects.  
\subsection{Seam Removal Ablation Study}\label{sec:ablation}
\begin{figure}[t]
    \centering
    \includegraphics[width=\columnwidth]{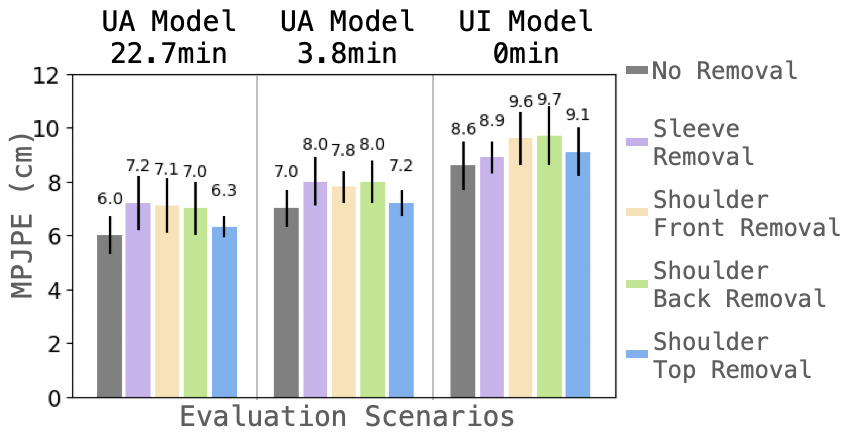}
    \caption{Seam Removal Ablation Study Results. A larger MPJPE indicates that the removed seam is more important. Error bars represent the standard deviation across participants.}
    \Description[]{This figure shows 3 bar graphs that display the Seam Removal Ablation Study results of: a User-Adaptive Model with 22.68 minutes of User Data, a User-Adaptive Model with 3.78 minutes of User Data, and a User-Independent Model with 0 minutes of User Data.The y-axis is the Mean Per Joint Position Error. The bars are color coded with gray being no seam removals, purple being sleeve removal, beige being shoulder front removal, green being shoulder back removal, and blue being shoulder top removal. For the user adaptive model with 22.7 min of training data, the user adaptive model with 3.78 min of training data, and the user-independent model with 0 min of user data, respectively, without any removals, the MPJPEs are 6.0, 7.0, and 8.6cm; with sleeve remove, the MPJPEs are 7.2, 8.0, and 8.9cm; with shoulder front removal, the MPJPEs are 7.1, 7.8, and 9.6cm; with shoulder back removal, the MPJPEs are 7.0, 8.0, and 9.7cm; with shoulder top removal, the MPJPEs are 6.3, 7.2, and 9.1 cm.}
    \label{fig:ablation-res}
\end{figure}

For our prototype, we chose a long-sleeve shirt with 8 seams, 4 each on the left and right side of the shirt.
Intrigued by the question of how different seam placements contribute to the model and how our approach will generalize with even fewer seams on a short-sleeve shirt and sleeveless shirt, we conducted a seam removal ablation study on the 6 odd-numbered participants.
Because the seam placements are symmetric, we remove 2 symmetric seams (\textit{e.g.}, left and right sleeve electrodes together) at a time, and train with the rest 6 seams.
We trained user-adaptive models and user-independent models for the 4 possible seam removals.
As shown in Fig.~\ref{fig:ablation-res}, with no surprise, given the small number of signal channels for a complex tracking task, removing any seam in any model leads to an increased error.
However, some seam placements have larger effects than others.

As we discussed in Sec.~\ref{sec:impl-signals}, the shoulder top seams have the smallest changes in magnitude, removing them does not harm the performance greatly: 0.2cm, 0.2cm, and 0.5cm worse with 22.7min, 3.8min, and 0min of user data.
The shoulder front/back seams have similar significance across all models, and removing them increases the errors by 1.0cm/0.9cm, 0.8cm/1.0cm, and 1.0cm/1.1cm with 22.7min, 3.8min, and 0min of user data.
On the other hand, the sleeve electrodes' importance varies across models.
For both adaptive models, the sleeve electrodes have the most important placements (\textit{i.e.}, their removal causes the largest errors compared with other placements) which lead to about 1cm additional error when removed.
For the user-independent model, the sleeve electrodes become the least important placement and only lead to an additional error of 0.3cm.
The observation implies that signals from some placements (sleeve) generalize across users worse than others do.
Furthermore, given the reasonable performance across all three models with the sleeve electrodes removed,~\theDevice{} has the potential to generalize on short-sleeve shirts and even sleeveless ones.

\subsection{Performance Comparison with Prior Works}
~\theDevice{} is the first to repurpose seams into capacitive sensors in clothing for body pose tracking.
Direct comparisons with other works are difficult because \theDevice{} and some other related works used different datasets to evaluate the performance. 
Factors like training data duration, pose/movement set design, data collection setup, etc. heavily affect the performance. Thus, we only include the comparison for an empirical comparison to help readers to situate our performance in the literature.  As this table \ref{tab:comparison} showed, ~\theDevice{} performance is comparable with prior wearable pose tracking systems while keeping the appearance of the shirt largely unchanged.
\begin{table}[t]
    \centering
    \caption{Situating~\theDevice{}'s tracking performance in the literature. ``UI'' stands for user-independent MPJPE.``UD'' stands for user-dependent/adaptive MPJPE.}
    \begin{tabular}{|c|c|c|c|c|c|}
        \hline
        &Tracking Device(s)& \begin{tabular}[c]{@{}c@{}}UI\\(cm)\end{tabular} &\begin{tabular}[c]{@{}c@{}}UD \\(cm)\end{tabular}\\
        \hline
        \begin{tabular}[c]{@{}c@{}}MI-Poser~\cite{MI-Poser}\\2023\end{tabular}
        & \begin{tabular}[c]{@{}c@{}}magnetic and inertial sensor \\ fusion on a headset\&2 controllers \end{tabular} & 6.6 & -\\
        \hline
        \begin{tabular}[c]{@{}c@{}}PoseSonic~\cite{PoseSonic}\\2023\end{tabular}
        & \begin{tabular}[c]{@{}c@{}}acoustic sensing \\ on smartglasses\end{tabular} & 6.2 & 5.6\\
        \hline
        \begin{tabular}[c]{@{}c@{}}LIP~\cite{LIP}\\2024\end{tabular}
        & \begin{tabular}[c]{@{}c@{}}4 IMUs \\ on a jacket\end{tabular} & 10.58 & -\\
        \hline
        \begin{tabular}[c]{@{}c@{}}MocaPose~\cite{Mocapose}\\2023\end{tabular}
         & \begin{tabular}[c]{@{}c@{}}16 conductive patches \\ on a jacket\end{tabular} & 8.8 & 8.6\\
        \hline
        \textbf{\theDevice{}}  & \begin{tabular}[c]{@{}c@{}}8 conductive seams \\ on a shirt\end{tabular} & 8.6 & 6.0\\
        \hline
    \end{tabular}
    \Description[]{This table situate SeamPose’s tracking performance in the literature. The table as 4 columns: the first column lists the project names; the second column lists the tracking device(s) in use; the third column lists the user-independent MPJPE in cm; the fourth column lists the user-dependent MPJPE in cm. MI-Poser, published in 2023, uses magnetic and inertial sensor fusion on a headset and 2 controllers and achieved a user-independent MPJPE of 6.6cm. PoseSonic, published in 2023, uses acoustic sensing on smartglasses and achieved a user-independent MPJPE of 6.2cm and a user-dependent MPJPE of 5.6cm. LIP, published in 2024, uses 4 IMUs on a jacket and achieved a user-independent MPJPE of 10.58cm. MoCaPose, published in 20243, uses 16 conductive patches on a jacket and achieved a user-independent MPJPE of 8.8cm and a user-dependent MPJPE of 8.6cm. SeamPose, our work, uses 8 conductive seams on a shirt and achieved a user-independent MPJPE of 8.6cm and a user-dependent MPJPE of 6.0cm.}
    \label{tab:comparison}
\end{table}

\subsection{Perceived Comfort}
In addition to its promising tracking ability, our sensing approach aims to preserve the soft and comfortable nature of clothing.
Based on the survey results, the participants felt very comfortable (Median=5 on the 5-point Likert scale; 1=very uncomfortable, 5=very comfortable) with the shirt prototype and perceived it as similar to their everyday clothing (Median=5 on the 5-point Likert scale; 1=very different, 5=not different at all).

\section{Discussion, Limitations \& Future Work}
~\theDevice{} demonstrates a minimally obtrusive continuous upper-body pose-tracking solution powered by a minimally altered shirt. The proof-of-concept prototype system was evaluated in a lab study to demonstrate its feasibility in tracking upper body pose.
In this section, we discuss the limitations of our current implementation and the opportunities and challenges of ubiquitous seam-enabled pose-tracking integration into everyday apparel.

\subsection{Improving Tracking Performance}
From prototyping the proof-of-concept shirt, we found the seam signal quality is mainly influenced by (a) the thread selection: conductivity and insulation (Sec.~\ref{sec:impl-fab}); and (b) the positions and number of seam electrodes (Sec.~\ref{sec:seam-patterns}).
We chose to constrain the latter for our design choice of minimal alteration and maximal generalizability and explored the positions and the number of seams with the seam removal ablation study, but an extensive characterization of such factors may improve the information gain.

Although our tracking performance is comparable with state-of-the-art loose clothing pose-tracking systems, we observed the drop in performance with unseen users and unseen movements (detailed in Sec.~\ref{sec:results}), a limitation that plagues many data-driven wearable sensing technologies.
Inspired by recent IMU-based pose-tracking advancements empowered by large synthetic datasets (generated from attaching virtual IMUs to SMPL meshes)~\cite{Imuposer,DIP, LIP}, we plan to explore simulating seam electrodes' capacitive measurements to address this issue. Prior works have simulated capacitive measurement on the human body for activity recognition~\cite{act-rec-simulation}, and more recently, Sch{\"o}ffmann used a 3D finite-element method to simulate capacitive touch readings for human-robot interaction~\cite{CapSense}. Additionally, other thread-based sensing approaches like triboelectric, piezoelectric, and impedance sensing~\cite{uKnit, NGs-motion,human-environment-knit} can be integrated into seams and become multi-model seam sensors to complement the current system and further improve the tracking performance. 

As shown in Fig.~\ref{fig:ft-curve}, adding more user-specific data to the current user-adaptive model seems to not significantly improve the performance. 
In the future, we plan to make prototypes of different sizes (\textit{i.e.}, small/medium/large) and introduce user descriptors (\textit{e.g.}, anthropometric data) as inputs to the deep learning pipeline to better learn the variations among different body shapes.
Moreover, more data from other users at a significantly larger scale for training the user-independent model may improve both the user-independent and -adaptive performances as shown in many prior systems.

Another limitation of our current implementation is that we do not estimate global orientation, even though relative joint tracking suffices for applications like activity recognition, rehabilitation, etc.
For use cases that require global orientation,~\theDevice{} could become a part of a multi-device pose tracking framework (\textit{e.g.}, along with headsets for hands-free AR/VR uses or smartphones for mobile interactions).

\subsection{Manufacturing at Scale for Everyday Wearing Experience}
In this paper, we repurposed the seams by machine-sewing conductive threads over existing seams. 
The fabrication process is simple and holds tremendous potential for integration into industrial processes.
If the insulated conductive thread (Sec.~\ref{sec:impl-fab}) can withstand mechanical stress induced by industrial sewing machines and/or sergers, not explored in this paper, the sensing seams can be easily integrated into the industrial scaled cut-and-sew process by simply replacing the bobbin thread with conductive thread.
Instead of repurposing existing seams like the method we propose in this paper, seams can be functional as soon as they are created at factories.

Beyond industrial mass manufacture, we share some challenges with current research related to garments when aiming for everyday uses, i.e., the garments need to be waterproof and wear-proof.
One key avenue is in designing a washable sensing shirt with a detachable or washable sensing board~\cite{textile-connector-review}.
Further, extensive characterization of durability and possible performance degradation over wears (\textit{e.g.}, hysteresis), and mitigation approaches are essential for real-world adoption.

\subsection{Evaluation in Real-World Settings}
We evaluated~\theDevice{} extensively in a controlled lab setting as a proof of concept to justify the feasibility of this approach. However, there are scenarios that our system needs to be further tested for real-world deployments. For instance, while the participants wore their own shirts (different materials with different dielectric constants and long-/short-/no sleeves with different amounts of skin contacts) underneath the shirt, we did not investigate putting clothes over our prototype, which is common when wearing a shirt. 

Moreover, our seam electrodes are capacitive sensors that inherently sense touches and proximity.
Because the seam electrodes are very small in size, they are very insensitive to changes in even near proximity~\cite{Mocapose,cap-gym-mat}.
At the signal level, there are no changes in the seam signals when the wearer touches electrically grounded and other objects that could cause electromagnetic (EM) interferences, but there are small (relative to body movements and direct electrode contacts) changes when other objects/people approach the electrodes within about 5cm.
This finding is consistent with prior work~\cite{Mocapose}.
As the tracking level, with our real-time pipeline, there is no obvious performance degradation in the above situations.
However, it remains rather sensitive to touches and direct contact, especially heavy ones causing thread deformation, which may happen during intense exercise or while sleeping. Therefore, to deploy the system for everyday pose-tracking wearing experiences, the system needs to be fine-tuned to be robust to these noises. We plan to take a data-driven approach to collect more data in these scenarios and update our machine-learning model to handle the noise as mentioned above.

\theDevice{} focuses on leveraging seams in a shirt for upper body motion estimation. Similar to prior work~\cite{Mocapose}, we were only able to evaluate our approach in one shirt with one kind of commonly seen seam placements. However, there are many different clothing patterns. For example, seam electrodes around the wrists at the sleeve cuffs could improve/enable forearm/palm orientation tracking, respectively. The seam pattern we chose has one of the smallest numbers of seams for long-sleeve garments, so we believe that ~\theDevice{} can be generalized to apparel with more seams, which offers more information on upper-body pose for the model to learn.  
\subsection{Different Seam Patterns in Everyday Apparel}

As the ablation study shows, the performance decreased as we removed the sleeve electrode but the model can still infer the upper body pose from the seam electrodes around the shoulders. This indicates our approach holds the potential to work on short-sleeve shirts and sleeveless shirts with a slightly worse performance. However, if the shoulder electrodes do not exist, such as on strapped and strapless tops, it is unclear how our system will work. In the future, we plan to integrate sensing seams into more clothing patterns, including tops and bottoms (\textit{i.e.}, pants, shorts, skirts).
Further, we chose the zigzag stitch as a common stitch for seaming on home sewing machines.
Zigzag stitches are stretchable and allow natural fabric stretches as the wearer moves. 
Stretching the fabric with sewn thread in any direction decreases the measured capacitances. 
After hundreds of sessions with our prototype, there is no performance degradation caused by permanent deformations. 
From our experience, the stitch type may impact the raw signal qualities but is unlikely to greatly impact the tracking performance.
Nevertheless, a formal characterization of stitch types (\textit{e.g.}, overlock stitches common for industrial seaming) remains essential for future work.

\section{Conclusion}
We presented~\theDevice{}, repurposing seams as capacitive sensors in a shirt for upper-body pose tracking. 
Without modifying the clothing surface,~\theDevice{} integrates motion-capturing capabilities into clothing unobtrusively and "invisibly".
In our proof-of-concept prototype, we machine-sewed conductive thread over existing seams in a long-sleeve shirt.
The capacitive readings of the seam electrodes change as the wearer's body pose changes.
Then the capacitive signals, acquired by an untethered sensing board, are processed by a customized deep-learning pipeline and continuously estimate 8 upper-body joint positions in 3D, relative to the pelvis.
We evaluated~\theDevice{} with a 12-participant user study and achieved a mean per joint position error (MPJPE) of 6.0 cm, comparable with that of related works, paving the way for everyday pose-tracking smart clothing.

\begin{acks}
This project was partially supported by the National Science Foundation
Grant Award No. 2239569.
We thank Owen Koonce for appearing in the video figure.
We also would like to thank the study participants and the reviewers.
\end{acks}

\bibliographystyle{ACM-Reference-Format}
\bibliography{main}

\appendix
\section{User Study Video Instruction Details}\label{sup:video-descriptions}
There are 54 instruction videos:
\begin{itemize}
    \item 1-20: 20 gestures from~\cite{Mocapaci} - lean forward, lean backward, lean to left, lean to right, turn left, turn right, shrug, pinch waist, forearm block, open arms, hands on the head, arms up, flappy bird, claps, walk (\cite{Mocapaci} indicates fake walk but we allowed the participant to locomote), butterfly swing, respect gesture, confuse gesture, frame picture, and stop gesture. 
    \item 21: golf swing
    \item 22: right tennis swing
    \item 23: left tennis swing
    \item 24: right basketball dribble
    \item 25: left basketball dribble
    \item 26: basketball shooting
    \item 27: punches with alternate hands
    \item 28: left arm swing on the side of the body (like walking)
    \item 29: right arm swing on the side of the body (like walking)
    \item 30: left arm swing in front of the body
    \item 31: right arm swing in front of the body
    \item 32: head sequence 1 - move the head downward and upward
    \item 33: head sequence 2 - rotate the head clockwise and counterclockwise
    \item 34: head sequence 2 - rotate the left and right
    \item 35: head sequence 3 - tilt the head from shoulder to shoulder
    \item 36: shoulder sequence 1 - move the shoulders up and down
    \item 37: shoulder sequence 2 - move the shoulders forward and backward
    \item 38: shoulder sequence 3 - rotate the shoulders forward and backward
    \item 39-41: arm sequence - with the arms down, the left/right/both arm(s) curl(s) from the inside, neutral, and outside tracks sequentially
    \item 42-44: arm sequence - with the arms open to the sides and parallel to the ground, the left/right/both arm(s) curl(s) from the inside, neutral, and outside tracks sequentially
    \item 45-47: arm sequence - with the arms front and parallel to the ground, the left/right/both arm(s) curl(s) from the inside, neutral, and outside tracks sequentially
    \item 48-50: arm sequence - with the arms raised over the head, left/right/both arm(s) curl(s) from the inside and neutral tracks sequentially (curing arms from the outside track are difficult to perform)
    \item 51-52: arm sequence - with the left/right arm open to the side, the right/left arm raises over the head and curl(s) from the inside and neutral tracks sequentially
    \item 53-54: arm sequence - with the left/right arm front and parallel to the ground, the right/left arm raises over the head and curl(s) from the inside and neutral tracks sequentially
\end{itemize}
\begin{table}[h]
    \centering
    \caption{Anthropometric data of participants.}
    \begin{tabular}{|c|c|c|c|c|c|}
        \hline
        &
        \begin{tabular}[c]{@{}c@{}}Arm Length \\ (cm)\end{tabular}& \begin{tabular}[c]{@{}c@{}}Bust \\ (cm)\end{tabular} & \begin{tabular}[c]{@{}c@{}}Waist \\ (cm)\end{tabular} &
        \begin{tabular}[c]{@{}c@{}}Height \\ (cm)\end{tabular} &
        \begin{tabular}[c]{@{}c@{}}Weight \\ (kg)\end{tabular} \\
        \hline
        mean & 56.6 & 89.8 & 76.8 & 169.7 & 63.3\\
        \hline
        std  & 3.3 & 6.3 & 7.7 & 10.6 & 11.7\\
        \hline
        max  & 65.0 & 99.0 & 92.0 & 190.5 & 83.9\\
        \hline
        min  & 53.0 & 78.0 & 66.0 & 153 & 49.9\\
        \hline
    \end{tabular}
    \label{tab:measurments}
\end{table}
\end{document}